\documentclass[journal]{IEEEtran}
\usepackage{amssymb}
\usepackage{cite}
\usepackage[cmex10]{amsmath}
\usepackage{graphicx}
\usepackage{amsmath}
\usepackage{epstopdf}
\usepackage{bm}
\usepackage[nooneline,flushleft]{caption2}
\usepackage{indentfirst,flushend}
\usepackage{color}
\usepackage{algorithm}
\usepackage{algorithmic}

\usepackage{setspace}
\usepackage{multirow}

\usepackage{amsfonts}

\begin{document}
\title{Cache-Enabled Coordinated Mobile Edge Network: Opportunities and Challenges}

\author{Shiwen~He,~\IEEEmembership{Member,~IEEE},
        Wei~Huang,~\IEEEmembership{Member,~IEEE},
        Jiaheng~Wang,~\IEEEmembership{Member,~IEEE},
        Ju~Ren,~\IEEEmembership{Member,~IEEE},
       Yongming~Huang,~\IEEEmembership{Senior Member,~IEEE},
       and~Yaoxue~Zhang~\IEEEmembership{Senior Member,~IEEE}
\thanks{S. He, J. Ren, and Y. Zhang are with the School of Computer Science and Engineering, Central South University, Changsha 410083, China. S. He is also with the Purple Mountain Laboratories, Nanjing 210096, China.(email: \{shiwen.he.hn, renju, zyx\}@csu.edu.cn). }
\thanks{W. Huang is with the School of Computer Science and Information Engineering, Hefei University of Technology, Hefei 230601, China. He is
also with the Anhui Province Key Laboratory of Industry Safety and Emergency Technology, Hefei, 230601, China (email: huangwei@hfut.edu.cn).}
\thanks{J. Wang and Y. Huang is the National Mobile Communications Research Laboratory, Southeast University, Nanjing 210096, China.(e-mail: \{jhwang, huangym\}@seu.edu.cn).}
\thanks{W. Huang, S. He, J. Wang, and Y. Huang are the corresponding authors.}
}
\maketitle
%%% Abstract.
\begin{abstract}
Cache-enabled coordinated mobile edge network is an emerging network architecture, wherein serving nodes located at the network edge have the capabilities of baseband signal processing and caching files at their local cache. The main goals of such an emerging network architecture are to alleviate the burden on fronthaul links, while achieving low-latency high rate content delivery, say on the order of the millisecond level end-to-end content delivery latency. Consequently, the application of delay-sensitive and content-aware has been executed in close proximity to end users. In this article, an outlook of research directions, challenges, and opportunities is provided and discussed in depth. We first introduce the cache-enabled coordinated mobile edge network architecture, and then discuss the key technical challenges and opening research issues that need to be addressed for this architecture. Then, to reduce the delivery latency and the burden on fronthaul links, new cache-enabled physical layer transmission schemes are discussed. Finally, artificial intelligence based cache-enabled communications are discussed as future research directions. Numerical studies show that several gains are achieved by caching popular content at the network edge with proper coordinated transmission schemes.
\end{abstract}

\IEEEpeerreviewmaketitle

\section{Introduction}

Driven by the visions of ultra-high-definition video, intelligent driving vehicles, and Internet of Things (IoTs), high data rate and low-latency content delivery is becoming an urgent problem for future mobile communication systems. In order to fulfill the demands of millisecond delivery latency and the cost efficiency of existing network deployments, the design of future radio access network (RAN) architectures needs to achieve a good balance between evolution and revolution.

Conventional cloud RANs (C-RANs) have driven a paradigm shift in operation and deployment of mobile communication networks. Aiming to the on-demand provisioning of resources, C-RAN features centralized resource management in the remote cloud with the capabilities of computation, control, and data storage. However, the latest developments of wireless communications have started to make a profound impact through massive connectivity between humans and computers, as well as via the massive proliferation of edge devices, which are associated with all surroundings devices. Meanwhile, the provision of various wireless services is also experiencing a fundamental change from the traditional communications (such as phone calls, e-mails, and web browsing) to the emerging high-rate communications with low content delivery latency (e.g., video streaming, push media, mobile applications download/updates, and mobile television (TV)). In those emerging scenarios, one of the key performance indications is the content delivery latency. The features of processing and collecting/distributing data in a centralized manner makes C-RANs difficult to satisfy the demands of emerging wireless traffics~\cite{8403948}.

\begin{figure}[!t]
\renewcommand{\captionfont}{\footnotesize}
\renewcommand*\captionlabeldelim{.}
\centering
\captionstyle{flushleft}
\onelinecaptionstrue
\centerline{\includegraphics[width=0.8\columnwidth,keepaspectratio]{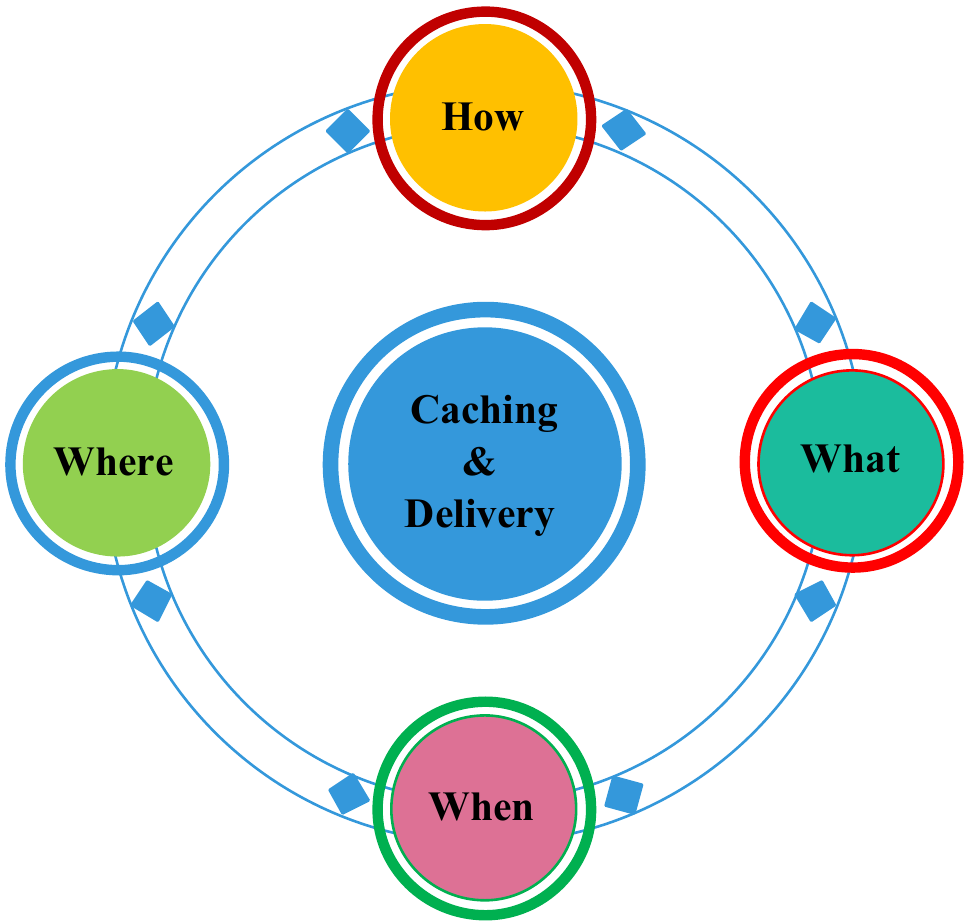}}
\caption{Key issues of cache-enabled wireless communication systems.}
\label{Edgecachechallengesfigure01}
\end{figure}

In view of the above issues related to C-RAN and the requirements of future communication scenarios, researchers in both academia and industry are investigating effective ways to satisfy the requirements of emerging traffics by adopting intelligent edge caching strategies. As a consequence, an emerging cache-enabled edge devices (EDs) coordinated mobile network architecture, called as ultra-dense cache-enabled cell-free (UD2CF) network, appears as a promising paradigm. This network has the ability to address the problems of low-latency and alleviate the large burden on the cloud processing units by deploying EDs with the capability of cache and baseband signal processing at the network edge. In the emerging UD2CF network, the EDs can pre-cache some popular files during the off-peak periods, even for the on-peak periods. Thus, in the on-peak periods, caching at the EDs can reduce both the delivery latency and the fronthaul burden and also relieve  the amount of mobile traffic during the on-peak periods. Furthermore, from the prior studies on cache-enabled wireless communication, ultra-dense edge nodes with the cache and signal processing functionalities have become apparent that the key technical issues are how, what, when, and where to cache or delivery issues, as illustrated in Fig.~\ref{Edgecachechallengesfigure01}.

\textbf {What and how to cache?} Cache at the network edge aims to achieve a trade-off between minimizing the transmission bandwidth cost, which is usually expensive, and reducing the storage cost, which is becoming much cheaper. Therefore, it is important to decide what files to cache by taking into account the file popularity. In general, only a few popular contents are frequently requested by a large portion of mobile users, while a majority of files are unpopular~\cite{5686876,6566245}. On the other hand, how to effectively place the popular contents at the network edge is also an important issue, which affects the gains of coordinated diversity and content distribution. In addition, caching strategies, i.e., deciding what and when to cache or update the cached contents, are crucial for the overall performance of cache-enabled communication systems. It is considerable important to estimate the gain behind a file by evaluating its current and potential popularity, storage size, and locations of existing replicas over the network topology.

\textbf {Where and when to cache?} Where should the frequently requested files be cached and when does one need to prefetch them to the network edge are the two key issues influencing the delivery latency and the overheads of fronthaul links. The conventional cache strategies pushing the requested content to the EN during off-peak periods become infeasible, because the popularity of contents and the network environments around the EDs as well as terminals are rapidly changing in the emerging UC2DF network. Therefore, such content and environment changes shall be taken into account in the study of the placement of the frequently requested files and timely cache them at the network edge, such that the delivery latency and the overheads of fronthaul links can be efficiently controlled or even reduced for the emerging UD2CF network.

In the above discussion, we mainly focus on the cache issues. In the following, we discuss the deliver issues and consider how the delivery latency can be further reduced. In particular, we will introduce the physical-layer and distributed transmissions, respectively.

\textbf {How to design physical-layer transmission?} How to fully exploit the capabilities of cache and baseband signal processing at the edge nodes to reduce the delivery latency and fronthaul link burden is a challenging issue for wireless transmission in the emerging UC2DF networks. An idea of hierarchy transmission that partitions the transmission into cache and network level transmission is demonstrated to be an effective way to achieve the above goal~\cite{8672586}.

\textbf{How to realize distribute transmission?} With an increasing number of EDs with the capabilities of cache and bandband signal processing, communication network is becoming much denser than ever before, which makes it much more difficult to build direct link among the EDs. This prohibits the signalling exchange among the EDs that may cause an unacceptable delay and signalling overhead. Therefore, each ED needs to perform the signal processing independently without signalling exchange. To this end, designing self-learning and self-optimization algorithm becomes a key issue to realize low-latency transmission and negligible signaling overhead among EDs.

\section{Emerging Network Architecture}

In the last few years, the heterogeneous cellular network and C-RAN are regarded as two most disruptive network architectures to improve the performance of wireless communication systems and satisfy the user demands with high quality of experience (QoE), such as high data rates, high spectral and energy efficiencies, and low-latency delivery, etc. Heterogenous cellular network provides high data rate transmission by deploying a large number of small base stations (SBS), while macro base stations (MBS) are designed to provide seamless coverage to a large area, which is an efficient approach to realize pathloss compensation and increase spectral and energy efficiency. Unfortunately, the ultra-dense SBS deployment may lead to extensive coverage overlaps among SBSs and serious inter-SBSs interference. C-RAN architecture, which decouples the radio frequency (RF) and baseband signal processing to the remote radio heads (RRHs) and baseband units (BBUs), respectively, has been proposed to improve the spectral and energy efficiency with more flexible resource exploitation. In the C-RAN architecture, the BBU is centralized as a BBU pool, which enables multi-points cooperative communications and also can efficiently suppress interference among the RRHs. However, the separation of the BBU and RF units makes the system performance depend on the fronthaul capacity that may results in inevitable delivery latency. In other words, the constraint of fronthaul links greatly impacts the system performance in the C-RAN architecture.

\begin{figure}[t]
\renewcommand{\captionfont}{\footnotesize}
\renewcommand*\captionlabeldelim{.}
\centering
\captionstyle{flushleft}
\onelinecaptionstrue
\includegraphics[width=1\columnwidth,keepaspectratio]{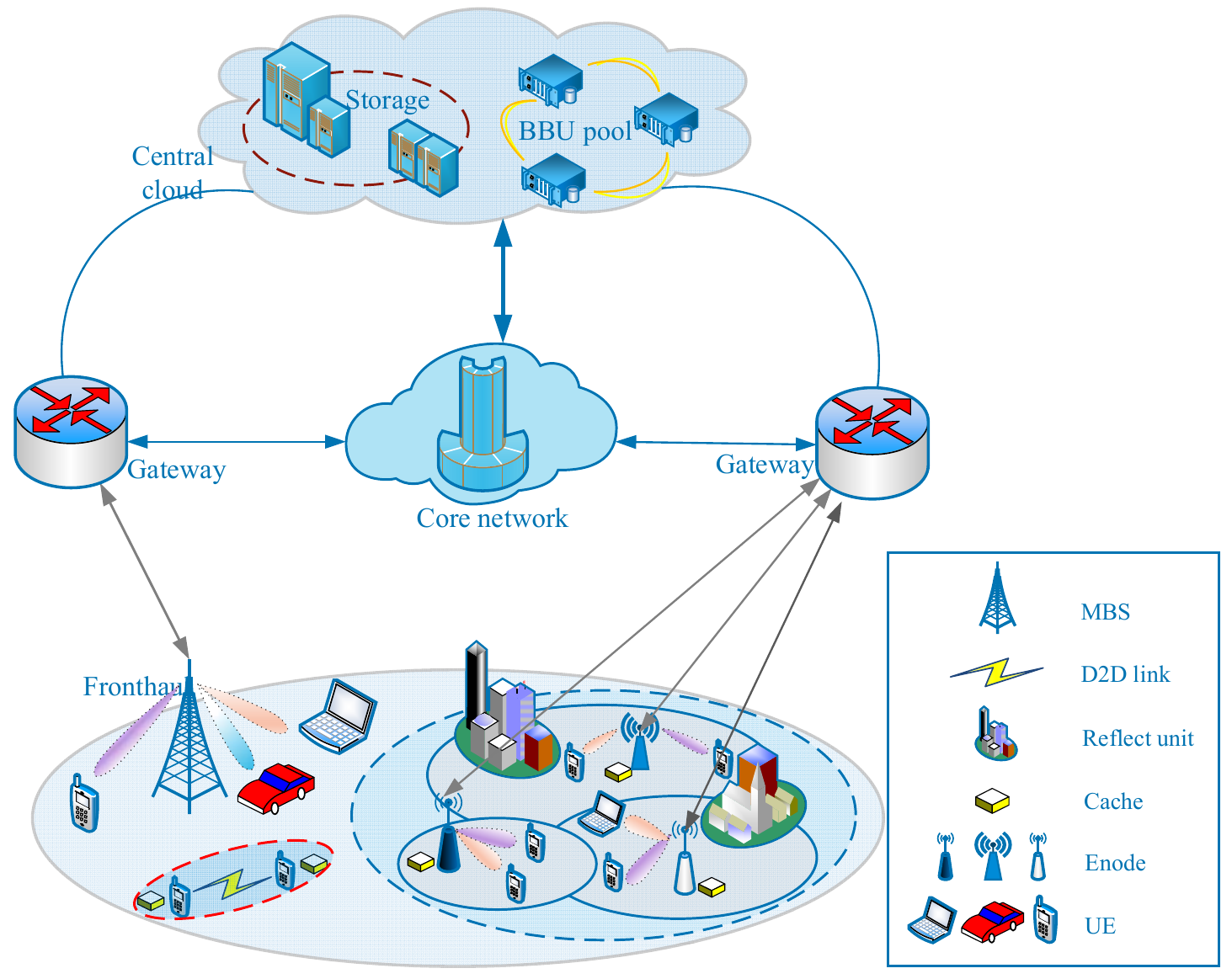}
\caption{Emerging cache-enabled ultra-dense network architecture.}
\label{NetworkArchitecture}
\end{figure}

With the emergence of new traffic types, end-to-end low-latency delivery (millisecond level) is imperatively needed by some specific scenarios, such as augmented reality (AR), virtual reality (VR), industrial control network, and vehicular communications. However, in the heterogeneous network and C-RAN network, the traffic data are usually stored at the remote data center, which leads to large delivery latency, especially for ultra-dense network. More recently, caching popular contents at the network edge becomes a powerful way to cater for the requirement of millisecond level delivery latency. As illustrated in Fig.~\ref{NetworkArchitecture}, the UD2CF network can be viewed as a hierarchical heterogeneous network, which is deployed at the edge of network infrastructure in order to fulfill the presented challenges. It is noted that the EDs are densely placed at the edge of the network, and thus can play a significant role in providing the seamless coverage, distributed cache, computing, baseband signal processing and networking infrastructure. Moreover, these EDs are connected to the remote cloud center via either the optical fiber or wireless fronthaul links, where the cloud center contains a BBU pool and storage equipments to realize the centralized processing.
Motivated by the new requirement of low-latency, the provision of various wireless services is experiencing a shift from the traditional connection centric, such as phone calls, e-mails, and web browsing, to the emerging content-centric communications. The key feature of these content-centric services is that a same content may be requested by multiple mobile users and the functionality of RRHs expands from the seamless coverage to advanced computer and cache. The RRH possessing the capabilities of cache and baseband signal processing is called as enhanced RRH (eRRH). In the emerging UD2CF architecture, the contents frequently requested by the users can be retrieved from the local cache or the remote data center, and the baseband signal processing can be performed at the eRRHs or BBUs in the remote cloud. The most frequently requested files can be pre-cached in the EDs during the off-peak traffic periods. In this way, the fronthaul overhead and delivery latency can be reduced during the peak traffic periods, even for the off-peak traffic periods. On the other hand, the BBUs located in the cloud center store all requested files, which facilitates the coordinated transmission among the EDs. As a cost, the coordinated transmission will consume the resource of the fronthaul links and incur latency. As such, the emerging network architecture needs to be carefully devised such that it can balance the capability of high capacity, low-latency, and large-scale access.

Though eRRHs have the potential capabilities of possessing more degree of freedom to retrieve locally data and providing baseband signal processing, to fully exploit the advantages of eRRHs at the network edge, there still exist some issues that need to be further explored for the UD2CF network, which are summarized as follows:
\begin{itemize}
\item Caching strategies: Due to the diversity of terminal behaviors and the limitation of cache capacity, how to efficiently cache the popular contents at the network edge becomes a primary problem for the UD2CF networks. In general, different caching strategies have different impacts on the system performance of UD2CF networks. On the other hand, caching popular contents during the off-peak or on-peak periods is another key problem.

\item Transmission strategies: To effectively reduce the delivery latency and alleviate the burden on fronthaul links, designing edge-cache based transmission scheme is urgent by using the cache and baseband signal processing functionality of eRRHs. On the other hand, aiming at the files uncached in the eRRHs, realizing low-latency delivery is also a crucial problem. Especially for the multiple users case, partial requested files are cached at the network edge while another requested files need to be fetched from the far data center.
\end{itemize}

\section{Cache Placement}

The UD2CF network architecture is regarded as a promising solution to significantly increase the transmission data rate, reduce the delivery latency, and alleviate the burden on fronthaul links. Caching the popular files at the network edge is the foundation of making the potential role of UD2CF network to achieve the goals. In general, finishing the requests of users consist of files pre-fetching and data delivery. In the pre-fetching phase, the objective is to pre-fetch the content to the eRRHs in a timely and reliable manner. Sequentially, the ideal situation in the data delivery phase is that all terminals requesting contents can be coordinately served by the eRRHs without communicating with the remote cloud center. It implies that the design of pre-fetch strategies under the finite cache size and the dynamic feature of wireless networks is critical for evaluating data delivery performance of UD2CF networks, because the unplanned cache at the eRRHs results in more inter-user or inter-cell interference. In other words, different caching strategies have different impacts on the system performance of UD2CF networks.

Usually, caching strategies are needed to consider the diversity of terminal behaviors as well as balance the transmission and content diversities. There are a large amount of literature focusing on this topics in the last few years. In the sequel, we list the four most frequently used caching policies:
\begin{itemize}
\item Random caching: All eRRHs randomly cache the files with equal probabilities regardless of the popularity distribution of those contents~\cite{8374917}.

\item Most popular caching: Each eRRH caches the most popular files until its memory space is full. For this caching strategy, the files cached in eRRHs are to be assigned the same proportion of cache space when the sizes of their memory space are the same. This strategy can enable cooperative transmission via multiple eRRHs based on the caching state of requested files, especially for that the popularity distribution of files is non-uniform. However, when the popularity of all files is equal distribution, the cache hit rate is low and the requested files are extracted from eRRHs or remote cloud center, which may cause the additional fronthaul overheads and transmission latency~\cite{8374917}.

\item Probabilistic caching: Different from the most popular caching strategy that each eRRH caches the same files, each eRRH caches a file randomly with certain probability according to the file popularity. The more popular the file is, the more likely it will be cached in each eRRH. The probabilistic caching strategy has a good balance between the cooperative transmission performance and hit rate~\cite{7488289}.

\item Hierarchical cooperative caching strategy: It is a promising solution to reduce the total access delay via cooperatively caching the user requested' data at the network edge EDs with limited storage space~\cite{Rodriguez2001Web}. The hierarchical cooperative caching strategy is to divide the cache space into three parts: self, friends, and strangers, which can balance selfishness (caching the content according to the user preference served by this eRRH) and unselfishness (assisting other eRRHs to cache). In the self part, the eRRHs cache the most popular content. For these eRRHs belonged to multiple small cells, they can be considered as friends to each other, and store the friends' popular content in their friends part. Meanwhile, each eRRH also randomly caches a subset of the remaining content into the strangers part. This caching strategy can efficiently reduce access delay and increase the cache hit probability and the future work will focus on the adaptive adjustment of the size of each part, according to the popularity of edge nodes in this network.
\end{itemize}

Nevertheless, performing caching only during off-peak period is not effective if the popularity of contents is rapidly changing or the cached files need to be frequently updated. Typical examples for this type of content include up-to-the-minute news, sports events requiring live updates, commerce promotion with frequent pricing changes, newly released music videos, etc. Non-orthogonal multiple access (NOMA)-assisted caching strategies with push-then-deliver and the push-and-deliver mode can realize the update of cache content during on-peak periods~\cite{8368286}. In addition, using machine learning approaches to explore and exploit the dynamic network environments to cache and update the caching contents at the network edge EDs is an effective and efficient way for the content-centric UD2CF networks.

\section{Physical Layer Transmission}

In the above section, we discuss some kinds of cache strategies under the UD2CF architecture. In order to achieve the low-latency goal, according to the communication procedures, some physical layer transmission schemes based on different cached strategies are provided in the following.

\subsection{Hierarchical Transmission}
In generally, for wireless cellular networks, the end-to-end delivery time is mainly rooted in the network layer, baseband signal processing, and propagation over the air interface. In the network layer, the data delivery between the eRRHs and BBU is usually going through routers. As a result, the greater the amount of data delivery, the larger the time delay, especially for multiple router network. On the other hand, if the fronthaul link adopts the wireless approach, the time overhead cannot also be ignored due to the contention of wireless channels and baseband signal processing. That is because the user scheduling and data queue still cause a certain time delay, especially for a network with a large number of users waiting for serves.

\begin{figure*}[!bbt]
\renewcommand{\captionfont}{\footnotesize}
\renewcommand*\captionlabeldelim{.}
\centering
%\captionstyle{flushleft}
\onelinecaptionstrue
\includegraphics[width=1.5\columnwidth,keepaspectratio]{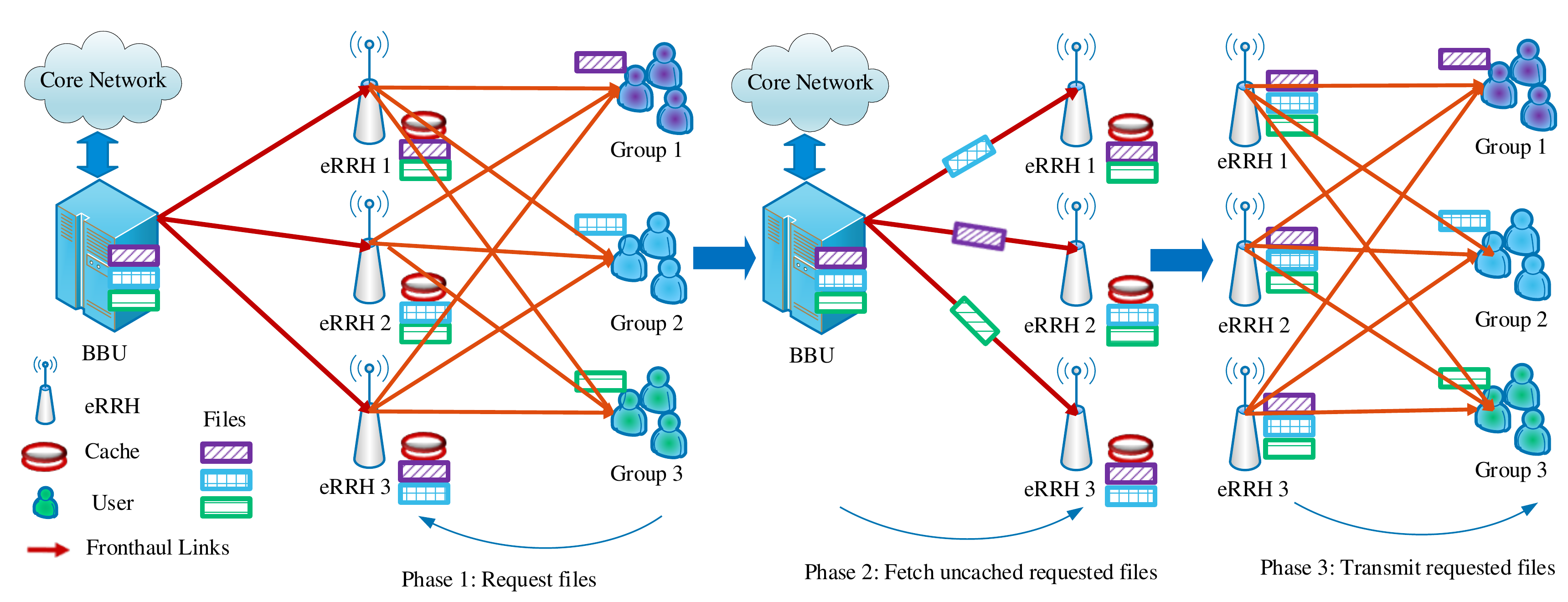}
%\centerline{\includegraphics[width=5.5in]{BulkTransmission.pdf}}
\caption{Flowchart of bulk transmission.}     \label{BulkTransmission}
\end{figure*}

In order to reduce the delivery latency, how to effectively and efficiently exploit the contents stored at the network edge and the data center to serve the requesting users become a key problem. In other words, edge-cache based physical layer transmission schemes need to be carefully design by using jointly the local cache and signal processing capabilities of eRRHs. Researchers in both industry and academia have done a lot of work on this issue. Design of dynamic content-centric eRRH clustering and beamforming with respect to both channel condition and caching status can effectively reduce the backhaul link overhead and the delivery delay~\cite{7488289}. Moreover, bulk transmission over the joint processing between the cloud and eRRHs can reduce the delivery latency for soft and/or hard transfer modes~\cite{7558153}, where eRRHs firstly pre-fetch the uncached requested files from the cloud, and then send both the cached and uncached requested files to the users, as illustrated in Fig.~\ref{BulkTransmission}. Though the bulk transmission can effective save the delivery latency by reducing the fronthaul overhead, waiting the coming of uncached requested files still waste a certain time. In other words, the cached and uncached requested files are delivered to users, simultaneously, which still spend much times for pre-fetching the uncached requested files.

\begin{figure*}[!bbt]
\renewcommand{\captionfont}{\footnotesize}
\renewcommand*\captionlabeldelim{.}
\centering
\onelinecaptionstrue
\includegraphics[width=2\columnwidth,keepaspectratio]{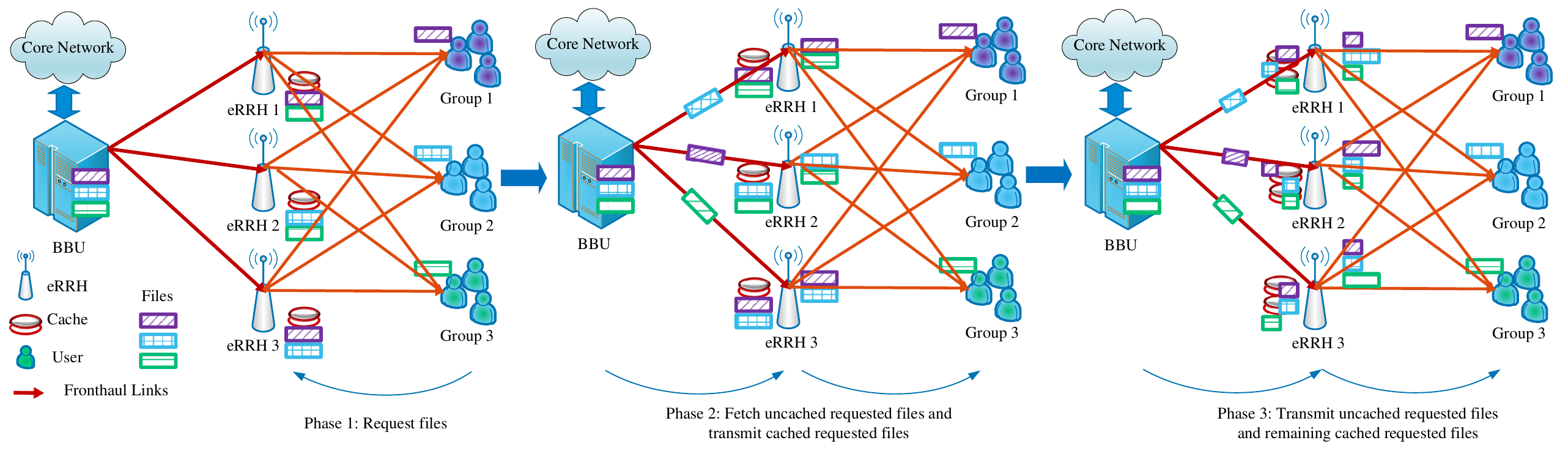}
\caption{Flowchart of pipelined transmission.}
\label{PipelinedTransmission}
\end{figure*}

To further reduce the delivery latency, a pipelined transmission scheme including cache-level and network-level transmission modes, is firstly proposed in~\cite{8463568,8672586}, as illustrated in Fig.~\ref{PipelinedTransmission}. This scheme consists of three phases, where the user firstly requests files in phase \uppercase\expandafter{\romannumeral1}. After receiving the requirements of users, in phase \uppercase\expandafter{\romannumeral2}, according to the caching status of requested files, the eRRHs transmit the cached requested files to the users while fetching the uncached requested files from the BBU. In phase \uppercase\expandafter{\romannumeral3}, after the arrival of uncached requested files, the eRRHs transmit the remaining cached requested files and uncached requested files to the users.

\begin{figure}[t]%[htbp]%here,top, bottom,page
\renewcommand{\captionfont}{\footnotesize}
\renewcommand*\captionlabeldelim{.}
\centering
\captionstyle{flushleft}
\onelinecaptionstrue
\includegraphics[width=0.8\columnwidth,keepaspectratio]{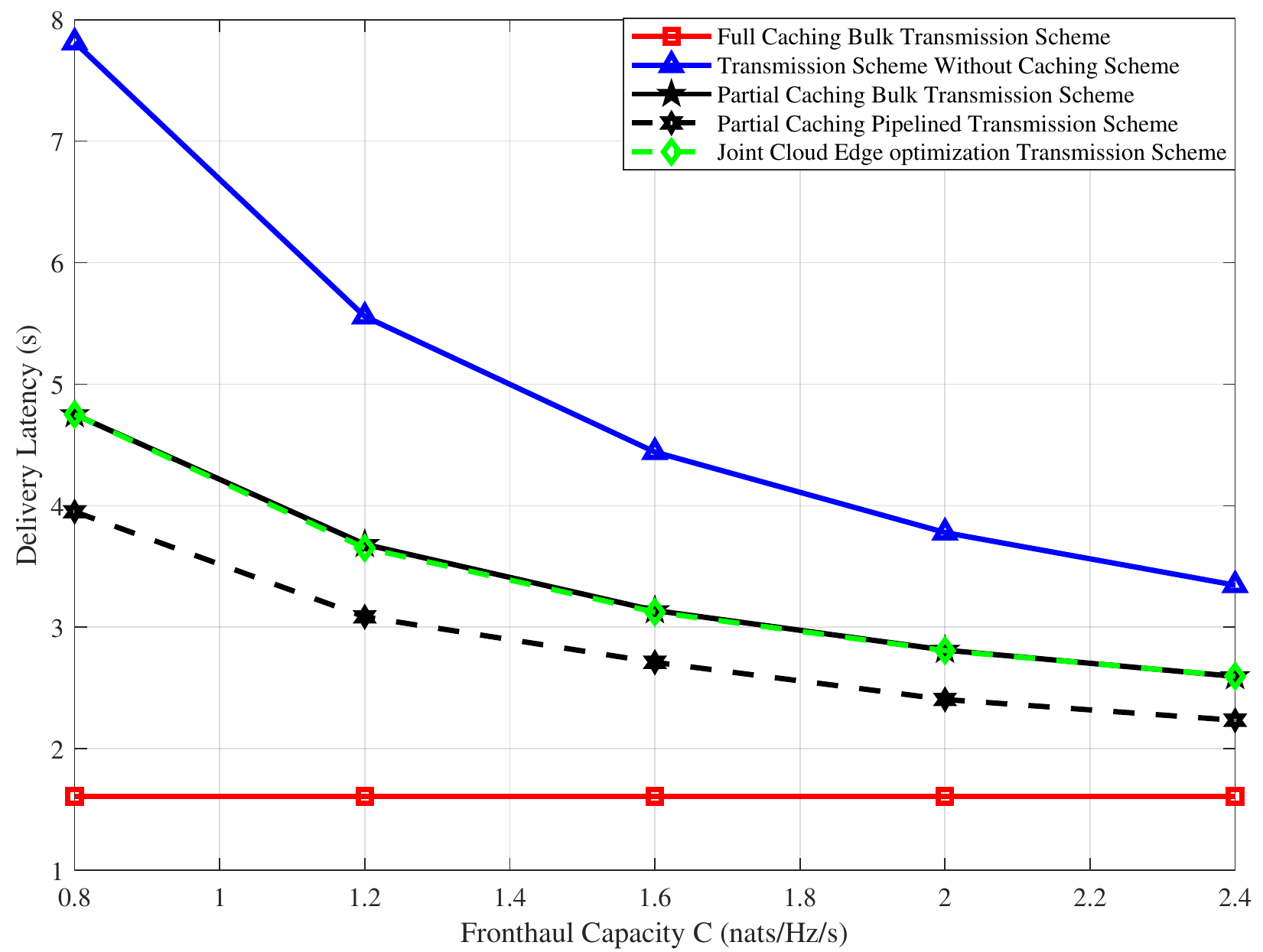}\\
\caption{Delivery latency versus fronthaul capacity $C$.}
\label{LatencyVsCapacity}
\end{figure}

The main difficult of realizing the pipelined transmission scheme is the synchronization between requested files at eRRHs. Fig.~\ref{LatencyVsCapacity} illustrates the delivery latency versus the capacity of fronthaul links for different transmission schemes, the details please see reference~\cite{8463568}. Numerical results reveal that exploiting fully the duration period consumed by fetching the uncached requested files to transmit cached requested files to users can further reduce the delivery latency. This is because the pipelined transmission scheme can provide more potential degree of freedom to coordinate the power allocation and manage the inter-user and inter-stream interferences.

Consequently, the hierarchical (bulk or pipeline) transmission schemes can effectively reduce the overhead of the fronthaul links and the delivery latency during the on-peak periods. However, in the UD2CF networks where EDs are ultra-densely deployed, the terminals usually are coordinately served by multiple EDs. In this case, the information exchange among the EDs will cause lots of time and signalling overheads. Therefore, decreasing the amount of the information exchange becomes a crucial issue, which will be discussed in the next subsection.

\subsection{Decentralized transmission}

In the UD2CF network, to perform full coordinated transmission among all EDs, the BBU must exchange the baseband signals with eRRHs and collect the global channel state information (CSI) from EDs to terminals. As the number of EDs and terminals increases, it becomes considerable difficult to collect the CSI such that full coordinated transmission cannot be realized. To overcome these difficulties, distributed and decentralized transmission schemes need to be designed to reduce the signalling overhead for coordinated communication systems. However, for the ultra-dense UD2CF network with demands of low-latency content delivery, the lack of direct link among EDs prohibits the signalling exchange (per-iteration) between coordinated EDs. This is because signalling exchange (per-iteration) between coordinated EDs no doubt causes a considerable delay and signalling overheads. Therefore, designing decentralized transmission scheme and self-organization approach is a promising way for avoiding signalling exchange among EDs in the UD2CF network~\cite{8454519}. Intuitively, if the ED can obtain all local CSIs, i.e., the channel coefficients between the ED to all terminals, a decentralized optimization algorithm can be independently implemented at each ED without any signalling exchange~\cite{8633840}. Furthermore, each ED may have the ability of reserving long-term and priori information in the communication processing, such as situational information, user behavior, historical data. Base on these reserved information and the enhanced baseband signal processing functionality of EDs, machine learning methods can be used to independently design novel decentralized transmission scheme, which achieves the goal of self-organizing communication manners and further reduces the transmission latency of whole network.

However, in the UD2CF network, the decentralized approaches can reduce the delivery latency caused by the information exchange among the EDs, but the delivery rate between EDs and terminals is still hard to satisfy the high data rate requirement of, e.g.,  augmented reality (AR) and virtual reality (VR) applications. Recently, millimeter wave (mmWave) communication emerges as an efficient transmission approach to increase the end-to-end delivery rate, due to the abundant bandwidth resources.

\subsection{Cache-enabled mmWave transmission}
The conventional communication systems operating in microwave frequency bands have almost reach their performance limits by fully digging the spatial-tempore-frequency resources. On the other hand, the appearances of various intelligent terminals require the UD2CF network to support higher data rate transmission.

To solve the scarcity of spectral resource and satisfy the demand on high data rate, mmWave communication owning abundant spectral resource is regarded as a promising solution for the UD2CF network. However, compared with Sub-6 GHz frequency bands, mmWave frequency bands encounter severe pathloss, penetration loss, and rain fading. Furthermore, they are easily absorbed or scattered by gases, especially for $60$ GHz frequency band that has attracted extensively studies in both academia and industry~\cite{7054722}. To this end, directional transmission via large-scale antenna array is regarded as a powerful method to compensate these shortcomings. On the other hand, directional transmission can efficiently suppress the inter-beam and inter-cell interferences.

From the perspective of interference management and network densification, mmWwave communication can not only solve the inter-cell and inter-user interference problem by exploiting directional transmission, but also effectively achieve high data rate transmission with the abundant available spectral resource in mmWave frequency bands. Therefore, the organic combination of UD2CF network and mmWave communication are expected to introduce amazing benefits. In other words, by exploiting the abundant spectral resource at mmWave frequency bands, the demands of high data rate (up to 20Gbps) can be easily achieved. Meanwhile, combining the network densification and directional transmission, the large path-loss of mmWave frequency bands can be easily compensated. It is very important to study the coordinated transmission mechanism, especially studying the decentralized coordinated mechanism, for cache-enabled mmWave UD2CF networks.

\section{Artificial intelligence (AI)-based technique in UD2CF}

\begin{figure}[!t]
\renewcommand{\captionfont}{\footnotesize}
\renewcommand*\captionlabeldelim{.}
\centering
\captionstyle{flushleft}
\onelinecaptionstrue
\centerline{\includegraphics[width=0.8\columnwidth,keepaspectratio]{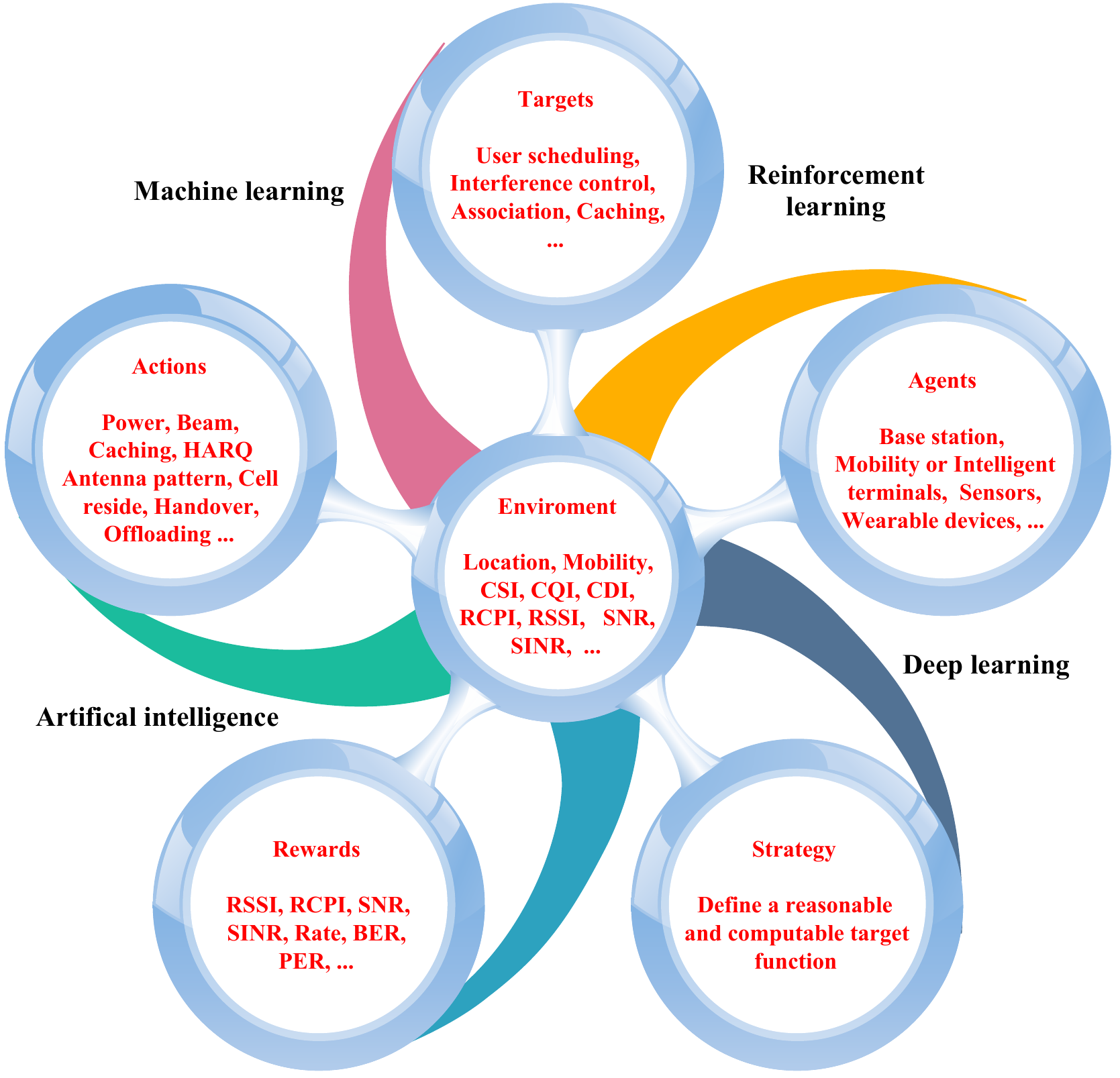}}
\caption{Artificial intelligence-based technique UD2CF architecture.}
\label{MachineLearning}
\end{figure}

For the complex dynamic environment of UD2CF networks, if the network operators continue to use conventional policies and methods, they may not be able to cope with the high-speed and low-latency demands due to the inter-user/inter-cell interference and massive user associations as well as the complicated resource allocation. Recently, artificial intelligence (AI) emerges as a promising tool to tackle problems encountered in caching, computing, and communications in wireless communication networks~\cite{8198798,8061008}. However, the application of AI for network traffic control and resource management, remains immature, due to the difficulty in uniquely characterizing the network environment and traffic features subject to an appropriate input and output dataset to the learning structures. The network environment and traffic features are anticipated to be even more dynamic and complex in UD2CF networks with high data rate and low-latency demands, which are coupled with actions, agents, rewards, and so on, as shown in Fig.~\ref{MachineLearning}.

In order to support dynamic network environment (such as: mobility, channel quality information (CQI), channel direction information (CDI), etc), the conventional approaches (action: power allocation, beam align, cache placement, handover.etc) need to be further optimized and updated for UD2CF networks. In other words, the research challenge consists of how to design an intelligent technique that can predict an environment action event a priori so that the strategy and agents can be changed accordingly to the potential action in a proactive manner. As a result, using the machine learning, deep learning, and reinforcement learning to attain the target (user scheduling, interference control, association, cache, etc) is an interesting and challenging work in the future.

\section{Conclusions}
It is generally agreed that current cellular communication networks cannot cope with the ever growing low-latency delivery demand of mobile users. The emerging UD2CF is a potential solution. In this article, we have provided an overview of the emerging UD2CF architecture from the latency demand perspective. Based on latency-driven evaluation, we have demonstrated that the pipelined transmission of in-network caching into network edge can potentially help reduce transmission latency. We also presented a number of promising research opportunities and relevant challenges, particularly related to caching strategy, distributed optimization, coordinated transmission, and cache-enabled mmWave communication. Conclusively, we have highlighted the roles that mmWave communication and AI approach can play in further improving the performance gains of UD2CF networks.

\bibliographystyle{IEEEtran}
\bibliography{len_bio}

\begin{thebibliography}{10}
\providecommand{\url}[1]{#1}
\csname url@samestyle\endcsname
\providecommand{\newblock}{\relax}
\providecommand{\bibinfo}[2]{#2}
\providecommand{\BIBentrySTDinterwordspacing}{\spaceskip=0pt\relax}
\providecommand{\BIBentryALTinterwordstretchfactor}{4}
\providecommand{\BIBentryALTinterwordspacing}{\spaceskip=\fontdimen2\font plus
\BIBentryALTinterwordstretchfactor\fontdimen3\font minus
  \fontdimen4\font\relax}
\providecommand{\BIBforeignlanguage}[2]{{%
\expandafter\ifx\csname l@#1\endcsname\relax
\typeout{** WARNING: IEEEtran.bst: No hyphenation pattern has been}%
\typeout{** loaded for the language `#1'. Using the pattern for}%
\typeout{** the default language instead.}%
\else
\language=\csname l@#1\endcsname
\fi
#2}}
\providecommand{\BIBdecl}{\relax}
\BIBdecl

\bibitem{8403948}
Z.~{Chang}, L.~{Lei}, Z.~{Zhou}, S.~{Mao}, and T.~{Ristaniemi}, ``Learn to
  cache: Machine learning for network edge caching in the big data era,''
  \emph{IEEE Wireless Commun.}, vol.~25, no.~3, pp. 28--35, June. 2018.

\bibitem{5686876}
G.~{Dan}, ``Cache-to-cache: Could isps cooperate to decrease peer-to-peer
  content distribution costs?'' \emph{IEEE Trans. Parallel and Distributed
  Sys.}, vol.~22, no.~9, pp. 1469--1482, Sep. 2011.

\bibitem{6566245}
B.~{Ramanan}, L.~{Drabeck}, M.~{Haner}, N.~{Nithi}, T.~{Klein}, and
  C.~{Sawkar}, ``Cacheability analysis of http traffic in an operational lte
  network,'' in \emph{2013 Wireless Telecommun. Sym. (WTS)}, Apr. 2013, pp.
  1--8.

\bibitem{8672586}
S.~{He}, J.~{Ren}, J.~{Wang}, Y.~{Huang}, Y.~{Zhang}, W.~{Zhuang}, and
  S.~{Shen}, ``Cloud-edge coordinated processing: Low-latency multicasting
  transmission,'' \emph{IEEE J. Sel. Areas in Commun.}, vol.~37, no.~5, pp.
  1144--1158, May. 2019.

\bibitem{8374917}
P.~{Sermpezis}, T.~{Giannakas}, T.~{Spyropoulos}, and L.~{Vigneri}, ``Soft
  cache hits: Improving performance through recommendation and delivery of
  related content,'' \emph{IEEE J. Sel. Areas in Commun.}, vol.~36, no.~6, pp.
  1300--1313, Jun. 2018.

\bibitem{7488289}
M.~{Tao}, E.~{Chen}, H.~{Zhou}, and W.~{Yu}, ``Content-centric sparse multicast
  beamforming for cache-enabled cloud {RAN},'' \emph{IEEE Trans. Wireless
  Commun.}, vol.~15, no.~9, pp. 6118--6131, Sep. 2016.

\bibitem{Rodriguez2001Web}
P.~{Rodriguez}, C.~{Spanner}, and E.~{Biersack}, ``Analysis of web caching
  architectures: {H}ierarchical and distributed caching,'' \emph{IEEE/ACM
  Trans. on Net.}, vol.~9, no.~4, pp. 404--418, Aug. 2001.

\bibitem{8368286}
Z.~{Ding}, P.~{Fan}, G.~{Karagiannidis}, R.~{Schober}, and H.~{Poor}, ``{NOMA}
  assisted wireless caching: Strategies and performance analysis,'' \emph{IEEE
  Trans. Commun.}, vol.~66, no.~10, pp. 4854--4876, Oct. 2018.

\bibitem{7558153}
S.~{Park}, O.~{Simeone}, and S.~{Shamai Shitz}, ``Joint optimization of cloud
  and edge processing for fog radio access networks,'' \emph{IEEE Trans. on
  Wireless Commun.}, vol.~15, no.~11, pp. 7621--7632, Nov. 2016.

\bibitem{8463568}
S.~{He}, C.~{Qi}, Y.~{Huang}, Q.~{Hou}, and A.~{Nallanathan}, ``Two-level
  transmission scheme for cache-enabled fog radio access networks,'' \emph{IEEE
  Trans. Commun.}, vol.~67, no.~1, pp. 445--456, Jan. 2019.

\bibitem{8454519}
B.~{Rong}, M.~{Dianati}, L.~{Zhou}, G.~{Karagiannidis}, and C.~{Wang}, ``5{G}
  mmwave small cell networks: Architecture, self-organization, and
  management,'' \emph{IEEE Wireless Commun.}, vol.~25, no.~4, pp. 8--9, Aug.
  2018.

\bibitem{8633840}
S.~{He}, Y.~{Wu}, J.~{Ren}, Y.~{Huang}, R.~{Schober}, and Y.~{Zhang}, ``Hybrid
  precoder design for cache-enabled millimeter wave radio access networks,''
  \emph{IEEE Trans. Wireless Commun.}, vol.~18, no.~3, pp. 1707--1722, Mar.
  2019.

\bibitem{7054722}
J.~{Zhang}, X.~{Huang}, V.~{Dyadyuk}, and Y.~{Guo}, ``Massive hybrid antenna
  array for millimeter-wave cellular communications,'' \emph{IEEE Wireless
  Commun.}, vol.~22, no.~1, pp. 79--87, Feb. 2015.

\bibitem{8198798}
Y.~{He}, F.~{Yu}, N.~{Zhao}, V.~{Leung}, and H.~{Yin}, ``Software-defined
  networks with mobile edge computing and caching for smart cities: A big data
  deep reinforcement learning approach,'' \emph{IEEE Commun. Mag.}, vol.~55,
  no.~12, pp. 31--37, Dec. 2017.

\bibitem{8061008}
Y.~{He}, N.~{Zhao}, and H.~{Yin}, ``Integrated networking, caching, and
  computing for connected vehicles: A deep reinforcement learning approach,''
  \emph{IEEE Trans. Veh. Tech.}, vol.~67, no.~1, pp. 44--55, Jan. 2018.

\end{thebibliography}

\end{document}